# Modified curved boundary scheme for two-phase lattice Boltzmann simulations

Y. Yu, Q. Li[*], and Z. X. Wen

*School of Energy Science and Engineering, Central South University, Changsha 410083, China*

*Corresponding author: qingli@csu.edu.cn

**Abstract**

The lattice Boltzmann (LB) method has gained much success in a variety of fields involving fluid flow and/or heat transfer. In this method, the bounce-back scheme is a popular boundary scheme for treating nonslip boundaries. However, this scheme leads to staircase-shaped boundaries for curved walls. Therefore many curved boundary schemes have been proposed, but mostly suffer from mass leakage at the curved boundaries. Several correction schemes have been suggested for simulating single-phase flows, but very few discussions or studies have been made for two-phase LB simulations with curved boundaries. In this paper, the performance of three well-known types of curved boundary schemes in two-phase LB simulations is investigated through modeling a droplet resting on a circular cylinder. For all of the investigated schemes, the results show that the simulated droplet rapidly "evaporates" under the nonslip and isothermal conditions, owing to the imbalance between the mass streamed out of the system by the outgoing distribution functions and the mass streamed into the system by the incoming distribution functions at each boundary node. Based on the numerical investigation, we formulate two modified mass-conservative curved boundary schemes for two-phase LB simulations. The accuracy of the modified curved boundary schemes and their capability of conserving mass in two-phase LB simulations are numerically demonstrated.

# 1. Introduction

In the past three decades, the lattice Boltzmann (LB) method (Aidun and Clausen, 2010; Chen and Doolen, 1998; He et al., 2019; Li et al., 2016a; Xu et al., 2017) has been developed into an efficient and powerful numerical approach for simulating fluid flow and heat transfer. Compared with traditional numerical methods, the LB method has some major advantages. First, in the LB method the fluid flow is described by a discrete kinetic equation based on the particle distribution functions, which is a more fundamental level than normal continuum approaches (Huang et al., 2015). Second, the fluid pressure and the strain tensor are available locally in the LB method, while the pressure field in traditional numerical methods is usually obtained by solving the Poisson equation. Moreover, the LB method is easy to parallelize because of its explicit algorithm and local interactions.

Unlike traditional numerical methods, which usually specify the macroscopic variables such as $\rho$ and $\mathbf{u}$ at the boundaries, the LB method applies the boundary condition to the particle distribution functions, giving more degrees of freedom (Krüger et al., 2017) than the set of macroscopic variables. The oldest and simplest boundary condition or scheme in the LB community is the bounce-back scheme, which assumes that, when a particle hits a solid wall, it is reflected back to where it originally came from. Besides its simplicity, the bounce-back scheme is also well-known for its advantage in conserving mass (Krüger et al., 2017). When the physical boundary nodes lay at the lattice nodes, the bounce-back scheme is called the standard bounce-back scheme. On the contrary, when the solid wall locates at the middle of the fluid boundary node and the node inside the solid, the scheme is referred to as the halfway bounce-back scheme (Guo and Shu, 2013).

The bounce-back scheme is usually applicable to straight walls and it leads to staircase-shaped boundaries for curved walls. Using a stair-shaped approximation, the fidelity of real geometry is lost and additional errors may be introduced (K. P. Sanjeevi et al., 2018). To improve the accuracy of LB simulations involving curved boundaries, various curved boundary schemes have been developed. The first attempt which considers the accurate shape of a curved solid wall was made by Filippova and Hanel (Filippova and Hänel, 1998). The basic idea of their scheme is to create ghost nodes inside the solid and then apply a linear interpolation to determine the unknown distribution functions of a boundary node. However, this scheme suffers from numerical instability when the boundary node is close to the solid wall and the relaxation time is close to 1. Later, Mei *et al*. (Mei et al., 1999) proposed an improved curved boundary scheme based on the work of Filippova and Hanel, which enhances the numerical stability. These two schemes have been demonstrated to be of second-order accuracy.

Subsequently, Bouzidi *et al*. (Bouzidi et al., 2001) devised a curved boundary scheme that does not require the interpolation from a ghost node inside the solid. The motivation of their scheme is to interpolate the distribution functions in the interior fluid region and to include additional information about the wall location during the bounce-back process. Nevertheless, they employed separate treatments for different regions. In order to avoid discontinuity in the boundary treatment, Yu *et al*. (Yu et al., 2003) developed a unified scheme of Bouzidi *et al*.'s algorithm. This type of boundary schemes is also second-order accurate and usually referred to as the interpolated bounce-back scheme. However, it introduces viscosity-dependence boundary slip. To solve this issue, Ginzburg and d'Humières (Ginzburg and d'Humières, 2003) proposed a multi-reflection boundary scheme, which can be viewed as an enhanced interpolated bounce-back scheme. The main feature of their scheme lies in that it determines the coefficients of the interpolation through a second-order Chapman-Enskog expansion on the interpolated distribution functions (Krüger et al., 2017).

Another type of curved boundary schemes involves only a single lattice node and is free of interpolation, which is often called the "one-point" or "single-node" boundary scheme. The advantage of this type of boundary schemes is its local computations. The first one-point curved boundary scheme was proposed by Junk and Yang (Junk and Yang, 2005). Nevertheless, this scheme needs to compute the inverse of a matrix at each boundary node. Recently, Zhao and Yong (Zhao and Yong, 2017) devised an alternative single-node boundary scheme. They considered a linear combination of the pre-collision and equilibrium distribution functions at the fluid boundary node and introduced five parameters, which were determined by the Maxwell iteration. Moreover, Tao *et al*. (Tao et al., 2018) recently developed a single-node curved boundary scheme for LB simulations of suspended particles. Some other developments of the boundary schemes for LB simulations involving curved walls can be found in a recent book written by Krüger *et al*. (Krüger et al., 2017).

Although various curved boundary schemes have been developed, they mostly suffer from the violation of mass conservation in LB simulations (Krüger et al., 2017), i.e., there will be mass leakage at each time step during the simulations. To enhance mass conservation, several improved curved boundary schemes have been proposed. Kao and Yang (Kao and Yang, 2008) presented an interpolation-free boundary scheme to improve the mass conservation of the boundary condition. Bao *et al*. (Bao et al., 2008) devised an improved version of Mei *et al*.'s boundary scheme by redefining the density term in the equilibrium distribution function of the ghost node in solid. In addition, Le Coupanec and Verschaeve (Le Coupanec and Verschaeve, 2011) developed a mass-conservative boundary scheme for LB simulations with tangentially moving walls. Recently, Sanjeevi *et al*. (K. P. Sanjeevi et al., 2018) found that the interpolation-free boundary scheme proposed by Kao and Yang (Kao and Yang, 2008) still leads to mass leakage.

To the best of the authors' knowledge, very few studies or discussions have been made in the literature regarding the mass conservation of two-phase LB simulations when applying the curved boundary schemes, although some curved schemes have been utilized in the LB studies of two-phase flows. Definitely, the halfway bounce-back scheme is capable of conserving mass in two-phase LB simulations, but it is inappropriate for certain two-phase flows, such as two-phase flows with phase change (Li et al., 2015; Li et al., 2018; Yu et al., 2018), e.g., boiling on curved surfaces (Kang, 2016). The stair-shaped approximation of the bounce-back scheme will introduce "artificial" nucleation sites and may cause significant errors for boiling simulations.

The present study is therefore devoted to investigating the performance of some well-known curved boundary schemes in simulating two-phase flows. Based on the numerical investigation, two modified mass-conservative boundary schemes are formulated for two-phase LB simulations with curved boundaries. The rest of the present paper is organized as follows. In Sec. 2 we introduce the LB method and three well-known types of curved boundary schemes. In Sec. 3 we investigate the performance of three types of curved boundary schemes in two-phase LB simulations through modeling a droplet resting on a circular cylinder. The modified mass-conservative curved boundary schemes are formulated in Sec. 4. Numerical validation and discussion are also given there. Finally, a brief summary is given in Sec. 5.

## 2. Lattice Boltzmann method and curved boundary schemes

### 2.1. The lattice Boltzmann method

The LB equation using the Bhatnagar-Gross-Krook (BGK) collision operator can be written as (Qian et al., 1992)

$$f_\alpha(\mathbf{x}+\mathbf{e}_\alpha\delta_t, t+\delta_t) - f_\alpha(\mathbf{x},t) = -\frac{1}{\tau}\left[f_\alpha(\mathbf{x},t) - f_\alpha^{eq}(\mathbf{x},t)\right] + \delta_t F_\alpha(\mathbf{x},t), \tag{1}$$

where $f_\alpha$ is the density distribution function, $f_\alpha^{eq}$ is the equilibrium density distribution function, $\mathbf{x}$ is the spatial position, $\mathbf{e}_\alpha$ is the discrete velocity in the $\alpha$ th direction, $t$ is the time, $\delta_t$ is the time step, $\tau$ is the non-dimensional relaxation time, and $F_\alpha$ is the forcing term. The space is usually discretized in such a way that $\mathbf{e}_\alpha \delta_t$ is the distance between two neighboring nodes. Hence after one time step, $f_\alpha(\mathbf{x},t)$ will arrive at its neighboring node along the lattice velocity direction $\mathbf{e}_\alpha$. Therefore the LB equation can be split into two processes: the "collision" process

$$f_\alpha^*(\mathbf{x},t) = f_\alpha(\mathbf{x},t) - \frac{1}{\tau}\left[f_\alpha(\mathbf{x},t) - f_\alpha^{eq}(\mathbf{x},t)\right] + \delta_t F_\alpha(\mathbf{x},t), \tag{2}$$

and the "streaming" process

$$f_\alpha(\mathbf{x}+\mathbf{e}_\alpha\delta_t, t+\delta_t) = f_\alpha^*(\mathbf{x},t), \tag{3}$$

where $f_\alpha^*(\mathbf{x},t)$ is the post-collision state of the density distribution function.

From Eqs. (2) and (3) it can be found that the collision process is completely local and the streaming process is completely linear. Using the two-dimensional nine-velocity (D2Q9) lattice model (Aidun and Clausen, 2010; Chen and Doolen, 1998; Li et al., 2016a; Qian et al., 1992), the equilibrium density distribution function is given by

$$f_\alpha^{eq} = \omega_\alpha \rho \left[1 + \frac{\mathbf{e}_\alpha \cdot \mathbf{u}}{c_s^2} + \frac{(\mathbf{e}_\alpha \cdot \mathbf{u})^2}{2c_s^4} - \frac{(\mathbf{u} \cdot \mathbf{u})}{2c_s^2}\right], \tag{4}$$

where $c_s = 1/\sqrt{3}$ is the lattice sound speed and the weights $\omega_\alpha$ are defined as $\omega_0 = 4/9$, $\omega_{1-4} = 1/9$, and $\omega_{5-8} = 1/36$. The kinematic viscosity is related to the non-dimensional relaxation time through $\nu = c_s^2 (\tau - 0.5) \delta_t$, and the macroscopic density and velocity are calculated by, respectively

$$\rho = \sum_\alpha f_\alpha, \quad \rho \mathbf{u} = \sum_\alpha f_\alpha \mathbf{e}_\alpha + \frac{\delta_t}{2} \mathbf{F}, \tag{5}$$

where $\mathbf{F}$ is the total force of the system and is incorporated into the LB equation through the forcing term $F_\alpha(\mathbf{x}, t)$ (Li et al., 2016a).

**2.2. Curved boundary schemes**

Figure 1 illustrates a curved wall together with its neighboring lattice nodes. The link between a fluid boundary node $\mathbf{x}_b$ and a solid node $\mathbf{x}_s$ is cut by the wall at point $\mathbf{x}_w$. The fraction of the intersected link in the fluid region is defined as (Guo and Shu, 2013; Krüger et al., 2017)

$$q = \frac{|\mathbf{x}_b - \mathbf{x}_w|}{|\mathbf{x}_b - \mathbf{x}_s|}, \tag{6}$$

where $0 < q \leq 1$ and $|\mathbf{x}_b - \mathbf{x}_s|$ depends on $\mathbf{e}_\alpha$.

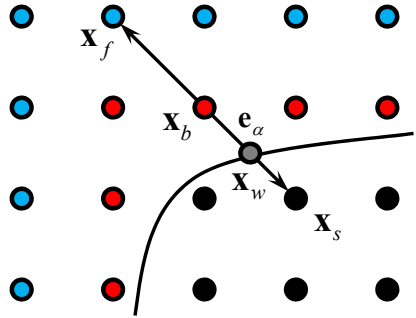

**Fig. 1**. Sketch of a curved wall and the lattice nodes near the wall.

The first LB boundary scheme for curved solid walls is proposed by Filippova and Hanel (Filippova and Hänel, 1998). In order to determine the unknown distribution function $f_{\bar{\alpha}}(\mathbf{x}_b, t + \delta_t)$ at the fluid boundary node (the subscript $\bar{\alpha}$ denotes the opposite direction of $\alpha$), Filippova and Hanel (Filippova and Hänel, 1998) defined a post-collision density distribution function at the solid node $\mathbf{x}_s$,

$$f_{\bar{\alpha}}^*(\mathbf{x}_s,t) = (1-\chi)f_\alpha^*(\mathbf{x}_b,t) + \chi f_\alpha^{eq}(\mathbf{x}_s,t). \tag{7}$$

The unknown distribution functions at the boundary node can be obtained by $f_{\bar{\alpha}}(\mathbf{x}_b,t+\delta_t) = f_{\bar{\alpha}}^*(\mathbf{x}_s,t)$ after the streaming process. The feature of this type of boundary schemes lies in that it introduces a fictitious equilibrium distribution function at the solid node, and Eq. (7) can be interpreted as a linear combination of the bounce-backed distribution function and the fictitious equilibrium distribution function, which is defined as

$$f_\alpha^{eq}(\mathbf{x}_s,t) = \omega_\alpha \rho(\mathbf{x}_b,t)\left[1 + \frac{\mathbf{e}_\alpha \cdot \mathbf{u}_s}{c_s^2} + \frac{(\mathbf{e}_\alpha \cdot \mathbf{u}_b)^2}{2c_s^4} - \frac{(\mathbf{u}_b \cdot \mathbf{u}_b)}{2c_s^2}\right], \tag{8}$$

where $\mathbf{u}_b = \mathbf{u}(\mathbf{x}_b,t)$ and $\mathbf{u}_s$ is a fictitious velocity. The parameter $\chi$ in Eq. (7) depends on the choice of the fictitious velocity $\mathbf{u}_s$. Filippova and Hanel (Filippova and Hänel, 1998) utilized the following choices:

$$q < 1/2: \quad \mathbf{u}_s = \mathbf{u}_b, \quad \chi = \frac{2q-1}{\tau-1}, \tag{9}$$

$$q \geq 1/2: \quad \mathbf{u}_s = \frac{q-1}{q}\mathbf{u}_b + \frac{1}{q}\mathbf{u}_w, \quad \chi = \frac{2q-1}{\tau}, \tag{10}$$

where $\mathbf{u}_w$ is the velocity at the point $\mathbf{x}_w$. In the present work, we only consider non-moving non-slip boundaries. Hence $\mathbf{u}_w = 0$. Obviously, when $q = 1/2$ ($\chi = 0$), the scheme of Filippova and Hanel reduces to the halfway bounce-back scheme, i.e., $f_{\bar{\alpha}}(\mathbf{x}_b,t+\delta_t) = f_\alpha^*(\mathbf{x}_b,t)$.

It can be found that the parameter $\chi$ will become very large when $\tau$ is close to 1 as $q < 1/2$, and then the scheme of Filippova and Hanel will suffer from severe numerical instability (Guo and Shu, 2013). To solve this problem, Mei *et al*. (Mei et al., 1999) proposed an improved scheme, which still uses Eq. (10) for $q \geq 1/2$ but employs the following choice for $q < 1/2$:

$$\mathbf{u}_s = \mathbf{u}_f, \quad \chi = \frac{2q-1}{\tau-2}, \tag{11}$$

where $\mathbf{u}_f = \mathbf{u}(\mathbf{x}_f,t)$ and $\mathbf{x}_f = \mathbf{x}_b - \mathbf{e}_\alpha \delta_t$ is the location of the fluid node beyond the boundary node (cf. Fig. 1). Mei *et al*. (Mei et al., 1999) showed that the above modification significantly improves the numerical stability of the boundary scheme.

The interpolated bounce-back scheme proposed by Bouzidi *et al*. (Bouzidi et al., 2001) employs an interpolation of the distribution functions in the interior fluid region. Using a linear interpolation, the boundary scheme of Bouzidi *et al*. is given by

$$f_{\bar{\alpha}}(\mathbf{x}_b, t+\delta_t) = \begin{cases} 2q f_\alpha^*(\mathbf{x}_b,t) + (1-2q) f_\alpha^*(\mathbf{x}_f,t), & q < \dfrac{1}{2} \\ \dfrac{1}{2q} f_\alpha^*(\mathbf{x}_b,t) + \dfrac{2q-1}{2q} f_{\bar{\alpha}}^*(\mathbf{x}_b,t), & q \geq \dfrac{1}{2} \end{cases}. \quad (12)$$

As $q = 1/2$, Eq. (12) reduces to the halfway bounce-back scheme. Alternatively, when a quadratic interpolation is used, the scheme takes the following form (Bouzidi et al., 2001):

$$f_{\bar{\alpha}}(\mathbf{x}_b, t+\delta_t) = \begin{cases} q(1+2q) f_\alpha^*(\mathbf{x}_b,t) + (1-4q^2) f_\alpha^*(\mathbf{x}_f,t) - q(1-2q) f_\alpha^*(\mathbf{x}_{ff},t), & q < \dfrac{1}{2} \\ \dfrac{1}{q(2q+1)} f_\alpha^*(\mathbf{x}_b,t) + \dfrac{2q-1}{q} f_{\bar{\alpha}}^*(\mathbf{x}_b,t) - \dfrac{2q-1}{2q+1} f_{\bar{\alpha}}^*(\mathbf{x}_f,t), & q \geq \dfrac{1}{2} \end{cases}, \quad (13)$$

where $\mathbf{x}_{ff} = \mathbf{x}_b - 2\mathbf{e}_\alpha \delta_t$. Some other versions of the interpolated bounce-back boundary scheme can be found in the book written by Krüger *et al*. (Krüger et al., 2017).

Another type of boundary schemes is the one-point or single-node boundary scheme. In comparison with the original single-node boundary scheme (Junk and Yang, 2005), Zhao and Yong (Zhao and Yong, 2017) recently devised a simpler single-node boundary scheme. They suggested a non-convex combination of $f_{\bar{\alpha}}^*(\mathbf{x}_b,t)$ and $f_\alpha(\mathbf{x}_b,t)$ at the fluid boundary node, which is defined as

$$f_{\bar{\alpha}}(\mathbf{x}_b, t+\delta_t) = \frac{2q}{1+2q} f_{\bar{\alpha}}^*(\mathbf{x}_b,t) + \frac{1}{1+2q} f_\alpha(\mathbf{x}_b,t). \quad (14)$$

In this scheme, the parameters/coefficients are determined based on an analysis using the Maxwell iteration. Obviously, Eq.(14) only involves local computations. In addition, it can be found that Eq. (14) does not reduce to the halfway bounce-back scheme when $q = 1/2$.

The scheme of Mei *et al*. (Mei et al., 1999), the scheme of Bouzidi *et al*. (Bouzidi et al., 2001), and the scheme proposed by Zhao and Yong (Zhao and Yong, 2017) can be regarded, respectively, as the representative schemes of the aforementioned three types of boundary schemes. In the remaining of this paper, these boundary schemes are referred to as the MLS scheme, the Bouzidi scheme, and the Zhao-Yong scheme, respectively. Moreover, we use "L-Bouzidi" and "Q-Bouzidi" to represent the Bouzidi scheme that utilizes the linear and quadratic interpolations, respectively.

# 3. Numerical investigation

## 3.1 Formulation of mass leakage

Figure 2 illustrates the distribution functions that will propagate across a curved wall through a boundary node (denoted by a solid black circle) during the streaming step. In the figure the blue arrows represent the outgoing distribution functions, which propagate from the boundary node to the ghost nodes in solid, while the red arrows represent the incoming distribution functions that propagate from the ghost nodes to the boundary node.

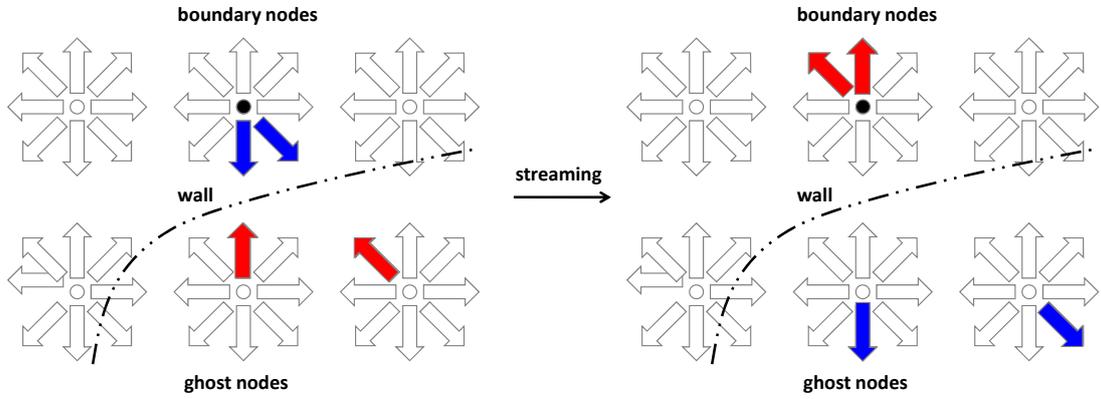

**Fig. 2**. Sketch of the distribution functions that propagate across a curved wall through a boundary node. The blue arrows represent the outgoing distribution functions, which propagate towards the ghost nodes in solid, while the red arrows represent the incoming distribution functions that propagate from the ghost nodes to the boundary node.

According to Fig. 2, we formulate the mass leakage at the boundary node as follows:

$$\Delta m(\mathbf{x}_b, t+\delta_t) = \sum_{\text{outgoing}} f_\alpha^*(\mathbf{x}_b, t) - \sum_{\text{incoming}} f_{\bar{\alpha}}(\mathbf{x}_b, t+\delta_t), \quad (15)$$

in which the first term on the right-hand side of Eq. (15) is the amount of mass streamed out of the system by the post-collision outgoing distribution functions while the second term on the right-hand side is the amount of mass streamed into the system by the incoming distribution functions. For the halfway bounce-back scheme, which gives $f_{\bar{\alpha}}(\mathbf{x}_b, t+\delta_t) = f_\alpha^*(\mathbf{x}_b, t)$, the mass leakage is zero. However, as previously mentioned, the halfway bounce-back scheme may cause considerable numerical errors for certain two-phase flows. For example, the stair-shaped approximation of the bounce-back scheme may introduce "artificial" nucleation sites for boiling or condensation on curved surfaces.

To illustrate the mass leakage caused by different boundary schemes, we firstly consider a test of single-phase flow, i.e., a static flow under gravity, which is illustrated in Fig. 3. The grid system is taken as $N_x \times N_y = 100 \times 50$ and the non-dimensional relaxation time $\tau$ is set to 1. The aforementioned three types of boundary schemes are employed at the upper and lower solid walls and the periodic boundary condition is applied in the *x* direction. The gravity force is given by $\mathbf{F}_g = (0, -\rho g)$, in which the gravitational acceleration is chosen as $g = 0.001$.

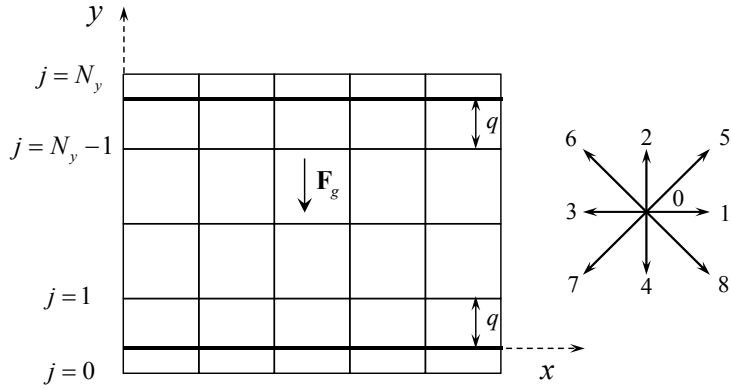

**Fig. 3**. Configuration of static flow under gravity with an arbitrary $q$.

In our simulations, the fraction $q$ defined by Eq. (6) is taken as $q = n/10$ with $n = 1, 2, \cdots, 9$. The variations of the normalized mass of the system are plotted in Figs. 4(a), 4(b), 4(c), and 4(d) for the MLS scheme, the L-Bouzidi scheme, the Q-Bouzidi scheme, and the Zhao-Yong scheme, respectively. As can be seen in these figures, the investigated boundary schemes basically exhibit different performance when different values of $q$ are applied. Specifically, for the MLS scheme, the L-Bouzidi scheme, and the Q-Bouzidi scheme, the mass of the system is conserved in the case of $q = 0.5$, but decreases in other cases, which is due to the fact that the MLS scheme and the Bouzidi scheme (regardless of using the linear or quadratic interpolation) reduce to the halfway bounce-back scheme in the case of $q = 0.5$.

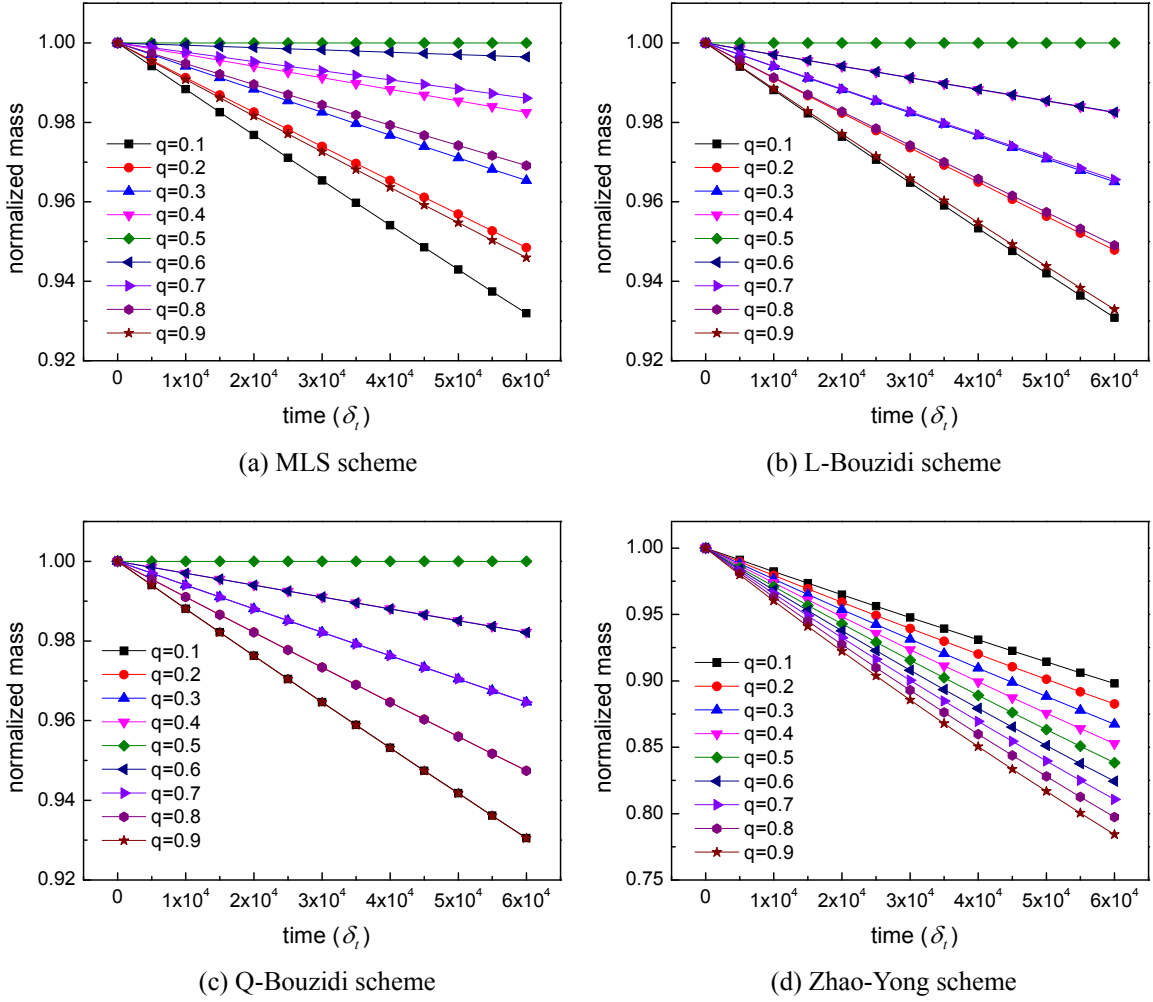

(a) MLS scheme  (b) L-Bouzidi scheme

(c) Q-Bouzidi scheme  (d) Zhao-Yong scheme

**Fig. 4**. Simulations of static flow under gravity using different boundary schemes. Variations of the normalized mass of the system against time.

To quantify the numerical results, an averaged mass change rate is evaluated at $t = 60000\delta_t$, which is defined as $\varepsilon = |m - m_0|/m_0 \times 100\%$. Here $m$ is the mass of the system at $t = 60000\delta_t$. The results are displayed in Fig. 5. From the figure we can clearly see that the mass change rate caused by the Zhao-Yong scheme is much larger than those produced by other schemes in all the cases. Such a finding indicates the single-node type of boundary schemes may perform worse than the other types of curved boundary schemes in terms of mass conservation although it has the advantage of local computations. Moreover, Fig. 5 also shows that there are no significant differences between the results of the MLS and Bouzidi schemes when $q < 0.5$, but the MLS scheme performs better as $q > 0.5$. In other words, the overall mass loss caused by the MLS scheme will be smaller than that caused by the Bouzidi scheme when the fraction $q$ in LB simulations varies from 0 to 1.

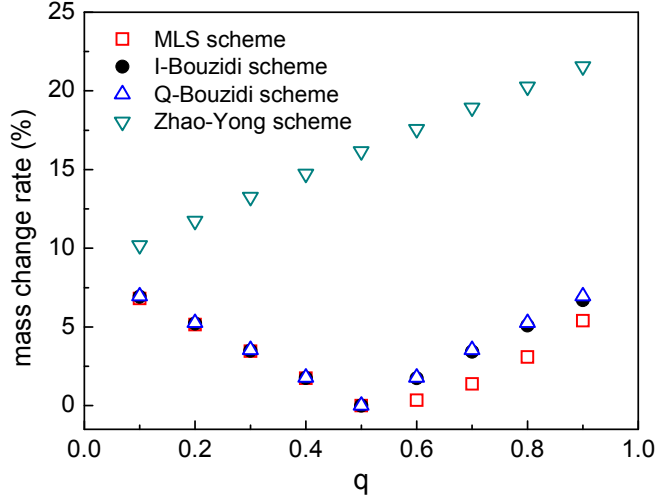

**Fig. 5**. Simulations of static flow under gravity using different boundary schemes. Variations of the averaged mass change rate against the fraction $q$ defined by Eq. (6).

**3.2. Simulations of droplet resting on a circular cylinder**

In this section, numerical simulations are carried out to investigate the performance of the MLS scheme, the Bouzidi scheme, and the Zhao-Yong scheme in two-phase LB simulations. The test of a droplet resting on a circular cylinder is employed. Under the non-slip and isothermal conditions, no evaporation will occur when a droplet is placed on a circular cylinder. However, if there exists mass leakage such as mass loss in LB simulations, the droplet will gradually become smaller, like an evaporating droplet. In our simulations, the isothermal pseudopotential multiphase LB method (Li et al., 2016a; Li et al., 2013; Shan and Chen, 1993, 1994; Xu et al., 2015; Zhang et al., 2014) is adopted, in which the fluid-fluid interaction force is defined as follows:

$$\mathbf{F} = -G\psi(\mathbf{x})\sum_{\alpha} w_{\alpha}\psi(\mathbf{x}+\mathbf{e}_{\alpha}\delta_{t})\mathbf{e}_{\alpha}, \quad (16)$$

where $\psi$ is the pseudopotential, $G$ is the interaction strength, and $w_{\alpha}$ are the weights. For the nearest-neighbor interactions on the D2Q9 lattice, the weights are given by $w_{\alpha}=1/3$ for $|\mathbf{e}_{\alpha}|^2=1$ and $w_{\alpha}=1/12$ for $|\mathbf{e}_{\alpha}|^2=2$.

The pseudopotential is chosen as $\psi(\rho)=\psi_0\exp(-\rho_0/\rho)$, where $\psi_0$ and $\rho_0$ are both set to 1.0,

which leads to the liquid density $\rho_L \approx 2.783$ and the vapor density $\rho_V \approx 0.3675$ when the interaction strength is taken as $G = 10/3$ (Li et al., 2016b). For this type of pseudopotentials, the forcing scheme proposed by Guo *et al*. (Guo et al., 2002) is employed to incorporate the fluid-fluid interaction force into the LB equation. The contact angle is implemented via a constant virtual wall density (Benzi et al., 2006; Huang et al., 2015; Li et al., 2014), which does not affect the conclusions in the present work. However, the contact angle scheme using a constant virtual wall density usually leads to an unphysical mass-transfer layer near the solid boundary. A recent study of the contact angle schemes involving curved boundaries can be found in a recent study of Li *et al*. (Li et al., 2019). In the present simulations, the computational domain is divided into $N_x \times N_y = 300 \times 350$ lattices. A circular cylinder of radius $R = 70$ is located at (150, 130) and a droplet is initially placed on the circular cylinder, as shown in Fig. 6. The periodic boundary condition is ultlized in the *x* and *y* directions and the kinematic viscosity is taken as $v = 0.15$ for both the liquid and vapor phases.

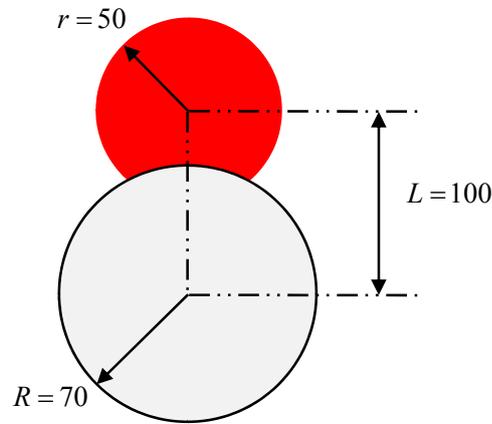

**Fig. 6** Initial setting of a droplet resting on a circular cylinder.

Some snapshots of the density contours obtained by the MLS scheme, the L-Bouzidi scheme, the Q-Bouzidi scheme, and the Zhao-Yong scheme are displayed in Figs. 7(a), 7(b), 7(c), and 7(d), respectively. From these figures it can be seen that, for all the investigated schemes, the simulated doplet gradually becomes smaller as time goes by, behaving like an evaporating droplet. However, the present simulations are performed using an isothermal pseudopotential LB model and the periodic boundary condition is applied in both the *x* and *y* directions, which means that there exists mass leakage at the surface of the circular cylinder due to the employed curved boundary schemes in simulations. Among these boundary schemes, the Zhao-Yong scheme has the advantage of local computations since it is a single-node scheme, while the other schemes usually require the information of two or more neighboring nodes. Similar to Fig. 4(d), Fig. 7(d) also shows that the Zhao-Yong scheme suffers from the violation of mass conservation although it does not involve any interpolation or extrapolation. Particularly, the rate of mass loss caused by the Zhao-Yong scheme is much larger than those caused by the other schemes, as shown in Fig. 8, which plots the variations of the normalized mass of the system during simulations. The normalized mass is defined as $\bar{m} = m/m_0$, where $m$ is the transient mass of the system and $m_0$ is the initial mass of the system at $t = 0$. From Fig. 8 we can also observe that the MLS scheme performs best among the investigated schemes, i.e., the mass of the system decreases relatively slowly in the test using the MLS scheme. Besides, it can be found that the performance of the Bouzidi scheme is better than that of the Zhao-Yong scheme, but worse than the MLS scheme.

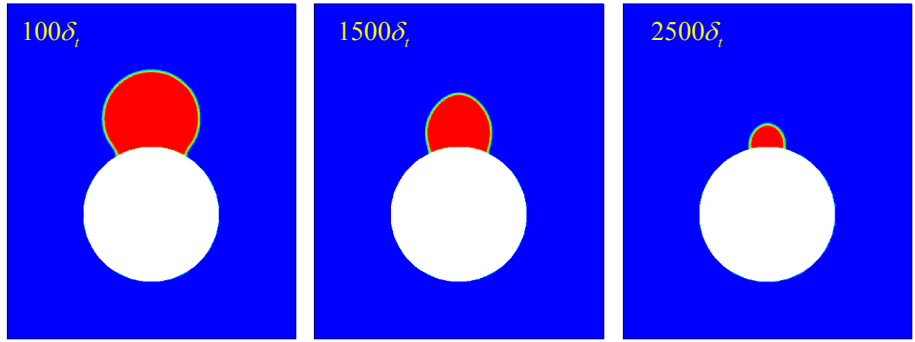

(a) MLS scheme

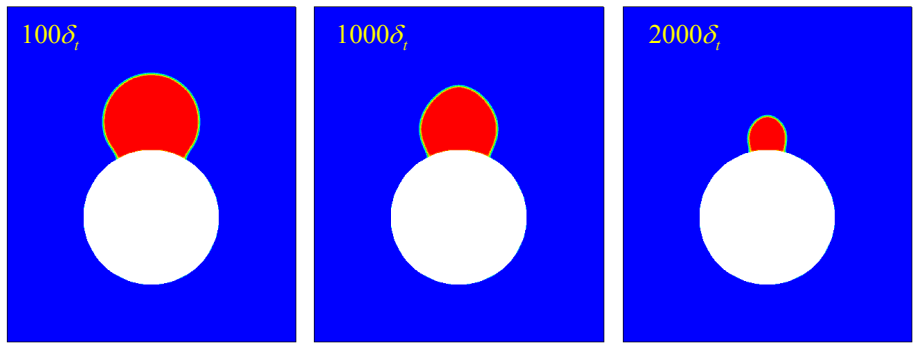

(b) L-Bouzidi scheme

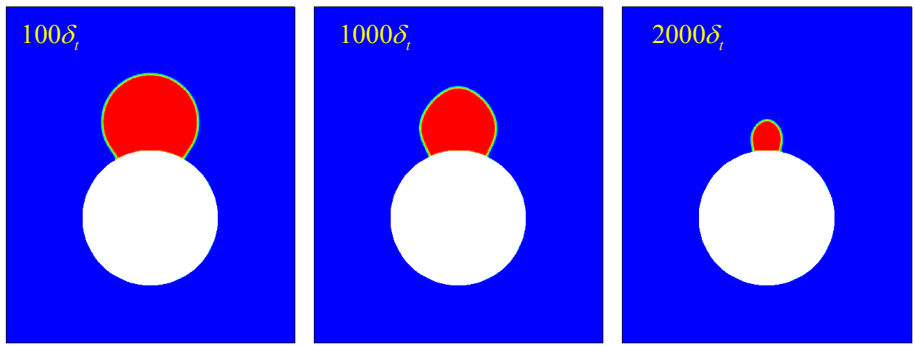

(c) Q-Bouzidi scheme

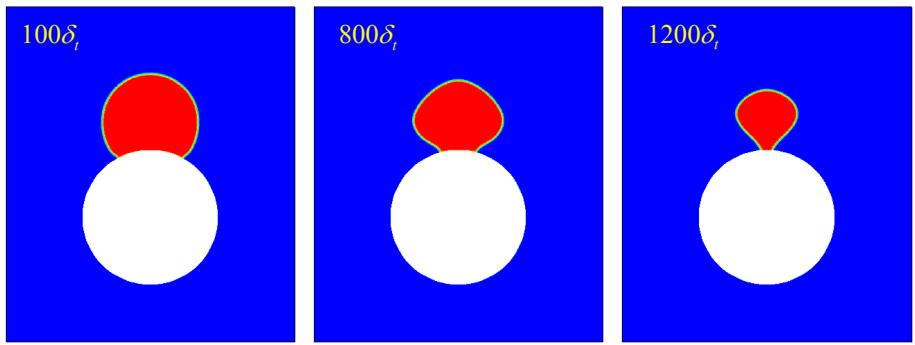

(d) Zhao-Yong scheme

**Fig. 7**. Snapshots of density contours obtained by different boundary schemes. (a) MLS scheme, (b) L-Bouzidi scheme, (c) Q-Bouzidi scheme, and (d) Zhao-Yong scheme.

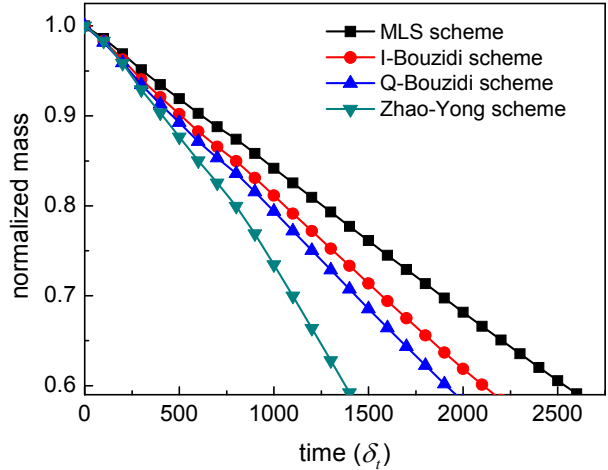

**Fig. 8**. Simulations of droplet on a circular cylinder using different boundary schemes. Variations of the normalized mass of the system against time.

By comparing the numerical results of two-phase LB simulations in this section with the results of single-phase LB simulations displayed in Sec. 3.1, we can observe that the mass in two-phase LB simulations decreases significantly faster than that in single-phase LB simulations. Taking the Zhao-Yong scheme as an example, in the present test it yields 40% mass loss after just 1500 time steps, as shown in Fig. 8. Such a rapid decrease of mass in two-phase LB simulations is probably due to the contact of the liquid-vapor interface with the curved solid boundary, which speeds up the mass leakage of the curved boundary schemes. When the curved boundary is completely immersed in liquid or vapor phase, the situation may be similar to that of single-phase LB simulations.

## 4. Modified mass-conservative schemes

### 4.1 Formulations of modified schemes

In this section, two modified mass-conservative curved boundary schemes are formulated. On the basis of the observation that the MLS scheme yields the smallest mass loss among the investigated curved boundary schemes, the modified boundary schemes are presented based on the MLS scheme. According to the MLS scheme, the unknown distribution functions at boundary nodes are determined by

$$f_{\bar{\alpha}}(\mathbf{x}_b, t+\delta_t) = (1-\chi) f_\alpha^*(\mathbf{x}_b, t) + \chi f_\alpha^{eq}(\mathbf{x}_s, t), \tag{17}$$

where $f_\alpha^{eq}(\mathbf{x}_s, t)$ is defined by Eq. (8) and the parameter $\chi$ is given by Eqs. (11) and (10) for $q < 0.5$ and $q \geq 0.5$, respectively. A simple treatment to conserve the mass of a system is to add the mass leakage back to the system. Such an idea was firstly used by Aidun and Lu (Aidun and Lu, 1995) for suspension flows when employing the halfway bounce-back scheme for moving boundaries. Similar treatments can also be found in (K. P. Sanjeevi et al., 2018; Liu et al., 2012). If the mass leakage is averagely added to the known distribution functions, the momentum at the boundary node may be changed. Hence the mass leakage given by Eq. (15) can be added to the rest distribution function and the following formula is obtained:

$$f_0(\mathbf{x}_b, t+\delta_t) = f_0^*(\mathbf{x}_b, t) + \sum_{\text{outgoing}} f_\alpha^*(\mathbf{x}_b, t) - \sum_{\text{incoming}} f_{\bar{\alpha}}(\mathbf{x}_b, t+\delta_t), \tag{18}$$

where $f_{\bar{\alpha}}(\mathbf{x}_b, t+\delta_t)$ are calculated by Eq. (17). In the remaining of the present paper, this modified MLS scheme is referred to as the modified scheme A. Such a modification is simple, but a possible limitation is also pointed out here, i.e., the rest distribution function $f_0(\mathbf{x}_b, t+\delta_t)$ may become negative in certain cases.

On the other hand, the mass of a system can also be conserved by guaranteeing the balance between the amount of mass carried by the outgoing and incoming distribution functions at each boundary node. When $\Delta m(\mathbf{x}_b, t+\delta_t)$ in Eq. (15) is set to zero, we can obtain

$$\sum_{\text{incoming}} f_{\bar{\alpha}}(\mathbf{x}_b, t+\delta_t) = \sum_{\text{outgoing}} f_\alpha^*(\mathbf{x}_b, t). \tag{19}$$

In order to satisfy the above requirement, an additional parameter is needed. Inspired by the study of Bao *et al*. (Bao et al., 2008), we can redefine the density in the fictitious equilibrium distribution function of the ghost nodes

$$f_\alpha^{eq}(\mathbf{x}_s, t) = \omega_\alpha \rho(\mathbf{x}_s, t)\left[1 + \frac{\mathbf{e}_\alpha \cdot \mathbf{u}_s}{c_s^2} + \frac{(\mathbf{e}_\alpha \cdot \mathbf{u}_b)^2}{2c_s^4} - \frac{(\mathbf{u}_b \cdot \mathbf{u}_b)}{2c_s^2}\right], \tag{20}$$

where the density has been changed to a fictitious density $\rho(\mathbf{x}_s, t)$. Substituting Eq. (17) into Eq. (19), the requirement given by Eq. (19) can be transformed to

$$\sum_{\text{outgoing}} \chi f_\alpha^*(\mathbf{x}_b, t) = \sum_{\text{outgoing}} \chi f_\alpha^{eq}(\mathbf{x}_s, t). \tag{21}$$

Combining Eq. (21) with Eq. (20), the fictitious density in Eq. (20) can be obtained as follows:

$$\rho(\mathbf{x}_s, t) = \frac{\sum_{\text{outgoing}} \chi f_\alpha^*(\mathbf{x}_b, t)}{\sum_{\text{outgoing}} \chi \omega_\alpha \left[1 + \frac{\mathbf{e}_\alpha \cdot \mathbf{u}_s}{c_s^2} + \frac{(\mathbf{e}_\alpha \cdot \mathbf{u}_b)^2}{2c_s^4} - \frac{(\mathbf{u}_b \cdot \mathbf{u}_b)}{2c_s^2}\right]}. \tag{22}$$

Since the parameter $\chi$ depends on $\mathbf{e}_\alpha$ through the fraction $q$, it cannot be moved out from the summation sign in Eqs. (21) and (22).

There is an important difference between the above formulation and that of Bao *et al*. (Bao et al., 2008). To obtain the fictitious density, Bao *et al*. assumed that the outing distribution functions, which are known at each boundary node, also satisfy Eq. (20) (such an assumption was illustrated above Eq. (12) in (Bao et al., 2008)). Then they obtained

$$\sum_{\text{outgoing}} f_\alpha^*(\mathbf{x}_b, t) = \sum_{\text{outgoing}} f_\alpha^{eq}(\mathbf{x}_s, t). \tag{23}$$

The above equation is just Eq. (12) in (Bao et al., 2008). Contrarily, our formulation given by Eq. (21) is directly derived from the requirement of Eq. (19), and we have emphasized that the parameter $\chi$ in Eq. (21) is dependent on $\mathbf{e}_\alpha$ through the fraction $q$. Hence $\chi$ in Eqs. (21) and (22) cannot be moved out from the summation over $\alpha$. In the following part, we will show that Eq. (23) still leads to mass leakage in two-phase LB simulations, although this equation seemingly works well for single-phase LB simulations in the study of Bao *et al*. (Bao et al., 2008). The modified MLS scheme based on Eqs. (20) and (22) is referred to as the modified scheme B.

### 4.2. Numerical validation

In this section, numerical simulations are performed to validate the two modified curved boundary schemes. Firstly, we employ a single-phase test, i.e., the test of steady flow past a circular cylinder, to examine the numerical accuracy of the modified schemes. The computational domain is chosen as $N_x \times N_y = 40D \times 25D$, where $D = 40$ is the diameter of the circular cylinder located at ($15D$, $12.5D$). The free-stream velocity is taken as $U_0 = 0.1$ and the Reynolds number is defined as $\mathrm{Re} = U_0 D/\nu$. The inlet boundary is given by a uniform velocity profile, which is implemented by the Zou-He boundary scheme (Zou and He, 1997), while the fully developed boundary condition is employed at the outlet boundary. The periodic boundary condition is applied to the upper and lower boundaries and the modified curved boundary schemes are used on the circular cylinder.

The non-dimensional length of the recirculation region, the separation angle, and the drag coefficient obtained by the modified schemes A and B are compared with some published results in Table 1. The non-dimensional length of the recirculation region is defined as $2L/D$, where $L$ is measured from the rearmost point of the cylinder. The drag coefficient is defined as $C_d = \bar{F}_x / (0.5 \rho U_0^2 D)$, where $\bar{F}_x$ is the *x*-component of the force $\bar{\mathbf{F}}$ exerted on the cylinder, which is calculated using the momentum-exchange method (Liao et al., 2015; Renwei et al., 2002). From the table we can see that the numerical results obtained by the modified schemes A and B are both in good agreement with those reported in the previous studies.

**Table 1**. Comparisons of the wake length, the separation angle, and the drag coefficient.

| References | Re = 20 | | | Re = 40 | | |
| --- | --- | --- | --- | --- | --- | --- |
| | $2L/D$ | $\theta_s$ | $C_d$ | $2L/D$ | $\theta_s$ | $C_d$ |
| Scheme A | 1.90 | 41.35 | 2.130 | 4.61 | 53.93 | 1.590 |
| Scheme B | 1.91 | 41.56 | 2.133 | 4.64 | 53.93 | 1.593 |
| (Dennis and Chang, 1970) | 1.88 | 43.7 | 2.045 | 4.69 | 53.8 | 1.522 |
| (Takeshi et al., 2014) | 1.90 | 40.78 | 2.104 | 4.64 | 50.38 | 1.568 |
| (Niu et al., 2006) | 1.89 | 42.95 | 2.144 | 4.52 | 53.86 | 1.589 |

Moreover, to illustrate the differences between the modified MLS schemes and the halfway bounce-back scheme, Fig. 9 presents a comparison of the local velocity vectors around the bottom of the circular cylinder at $Re = 40$ between the results of the modified schemes and those of the halfway bounce-back scheme. As shown in the figure, the halfway bounce-back scheme leads to non-physical results below the circular cylinder as the velocity obtained by the stair-shaped approximation penetrates the real cylindrical surface, which there is no such a phenomenon in the numerical results of the modified schemes A and B.

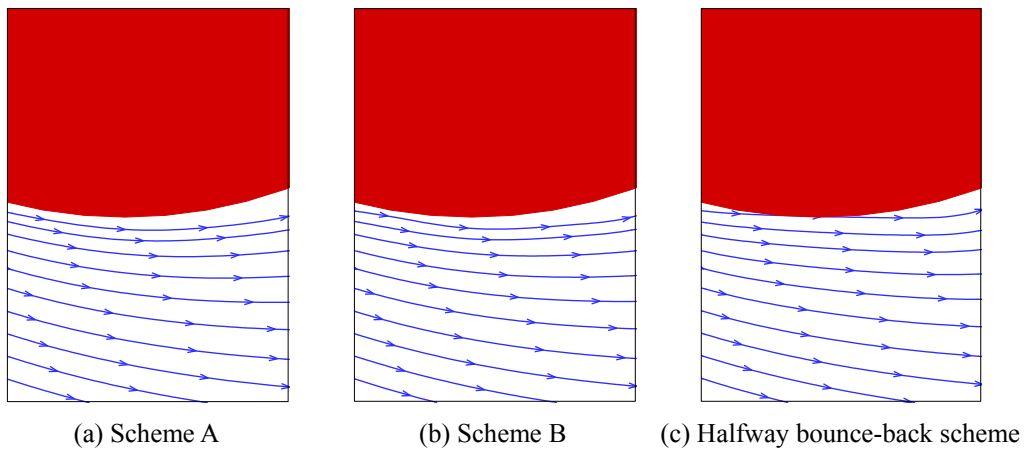

(a) Scheme A  (b) Scheme B  (c) Halfway bounce-back scheme

**Fig. 9**. Comparison of local velocity vectors around the bottom of the circular cylinder at $Re = 40$ between the modified curved boundary schemes and the halfway bounce-back scheme.

Now we turn our attention to the capability of the modified curved boundary schemes in conserving mass in two-phase LB simulations. The test of a droplet placed on a circular cylinder is still employed and the initial setting can be found in Fig. 6. Three cases with different contact angles are considered, i.e., $\theta = 30°$, $90°$, and $120°$. The grid system and the simulation parameters are the same as those used in Sec. 3.2. Some snapshots of the density contours obtained by the modified scheme A for the three cases are displayed in Fig. 10. By comparing the numerical resutls in Fig. 10 and those in Fig. 7, we can see that the droplet can achieve a stable state when the modified scheme A is employed, rather than behaving like an evaporating droplet when the original MLS scheme is used, which indicates that the modification formulated by Eq. (18) has prevented the mass leakage at the surface of the circular cylinder.

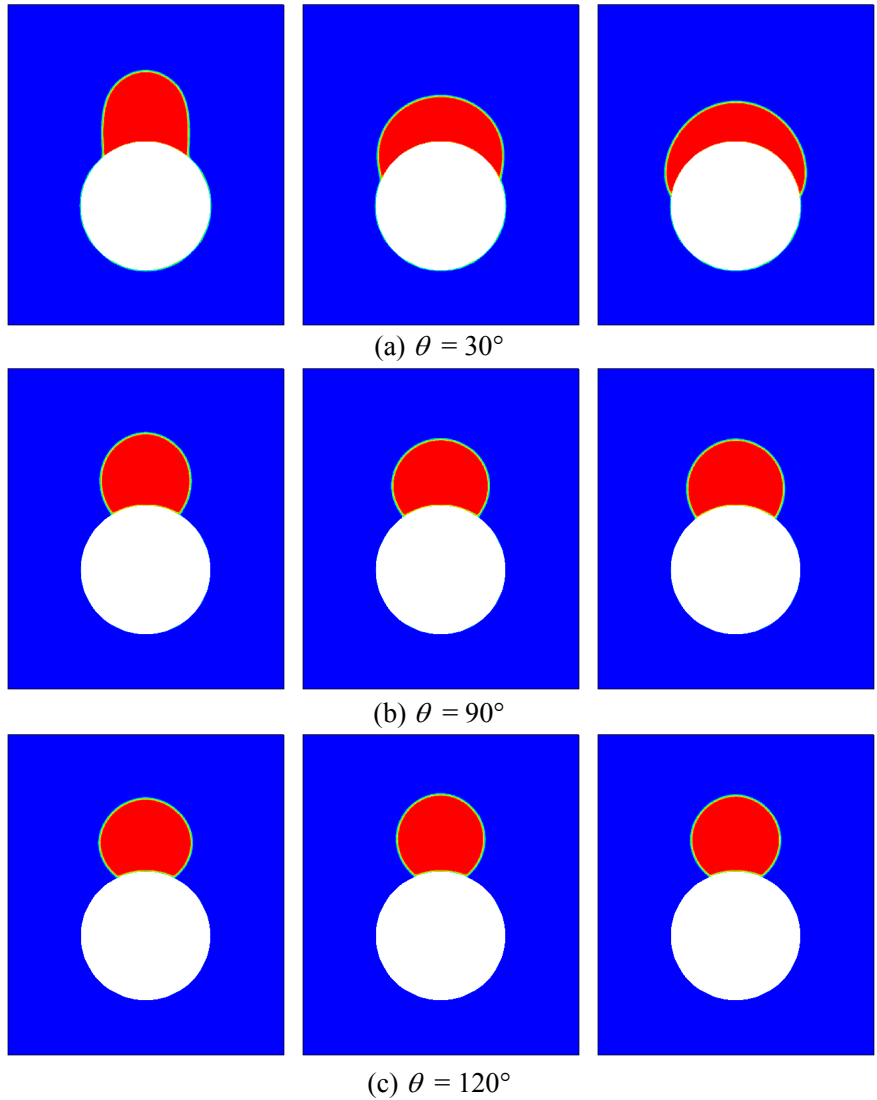

(a) $\theta = 30°$

(b) $\theta = 90°$

(c) $\theta = 120°$

**Fig. 10**. Simulations of droplet on a circular cylinder using the modified curved boundary scheme A. Snapshots of density contours under the static contact angles $\theta = 30°$, $90°$, and $120°$, respectively.

From left to right: $t = 1000\delta_t$, $5000\delta_t$, $20000\delta_t$.

Figure 11 displays the variations of the normalized mass of the system during the simulations using the modified schemes A and B, respectively. For both modified schemes, the maximum variation of the normalized mass of the system is very small (about 0.01%) and it appears at the early stage of the simulations, which is expected because the liquid-vapor interface is initially a sharp interface and it develops into a diffuse interface with a finite thickness at the early stage of the simulations. Such a change causes a small fluctuation of the mass of the system. After the early stage, the variations of the normalized mass of the system are much smaller, as can be seen in Fig. 11. In addition, we can also observe some slight differences between the results of different cases ($\theta$ = 30°, 90°, and 120°), which are attributed to the fact that the perimeter/length of the liquid-vapor interface is slightly different in these cases. To sum up, it is well demonstrated that both the modified scheme A and scheme B are capable of conserving mass in two-phase LB simulations, although they are devised based on the MLS boundary scheme from different points of view.

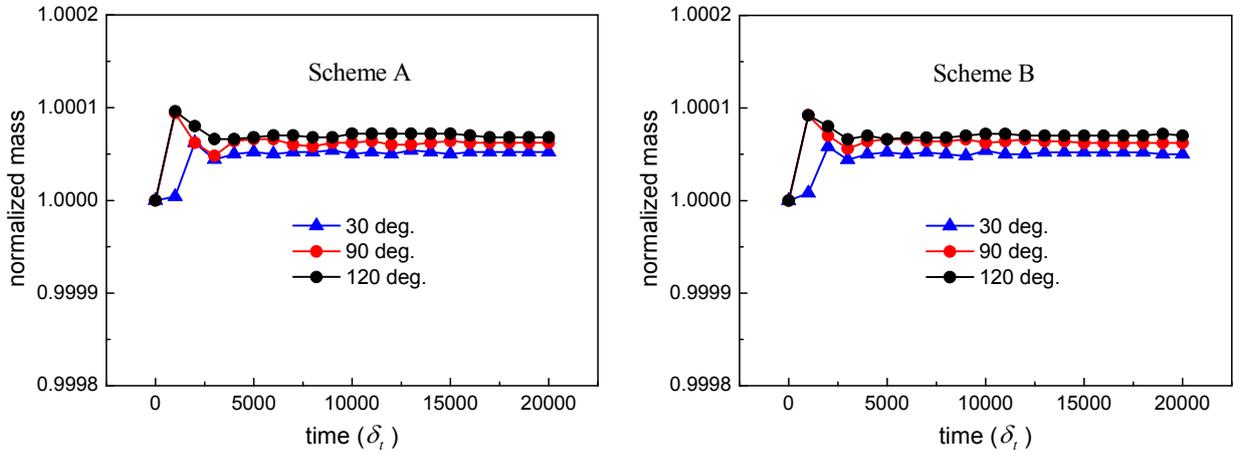

**Fig. 11**. Variations of the normalized mass of the system under the static contact angles $\theta$ = 30°, 90°, and 120°, respectively.

A comparison of the modified scheme B and the scheme of Bao *et al*. (Bao et al., 2008) is made in Fig. 12 by taking the case of $\theta$ = 90° as an example. From the figure it can be clearly seen that the scheme of Bao *et al*. still results in mass leakage in two-phase LB simulations although it performs much better than the original MLS scheme (see Fig. 7a). The reason for such a phenomenon has been previously mentioned, i.e., the fictitious density in the scheme of Bao *et al*. (Bao et al., 2008) was obtained by assuming that the outgoing distribution functions, which are known at each boundary node, satisfy Eq. (20). Such an assumption leads to the aforementioned Eq. (23), which is equivalent to moving the parameter $\chi$ (dependant on $\mathbf{e}_\alpha$) in Eq. (21) out from the summation over $\alpha$. Accordingly, it suffers from the violation of mass conservation, although such a defect was not detected by the single-phase LB simulations in (Bao et al., 2008).

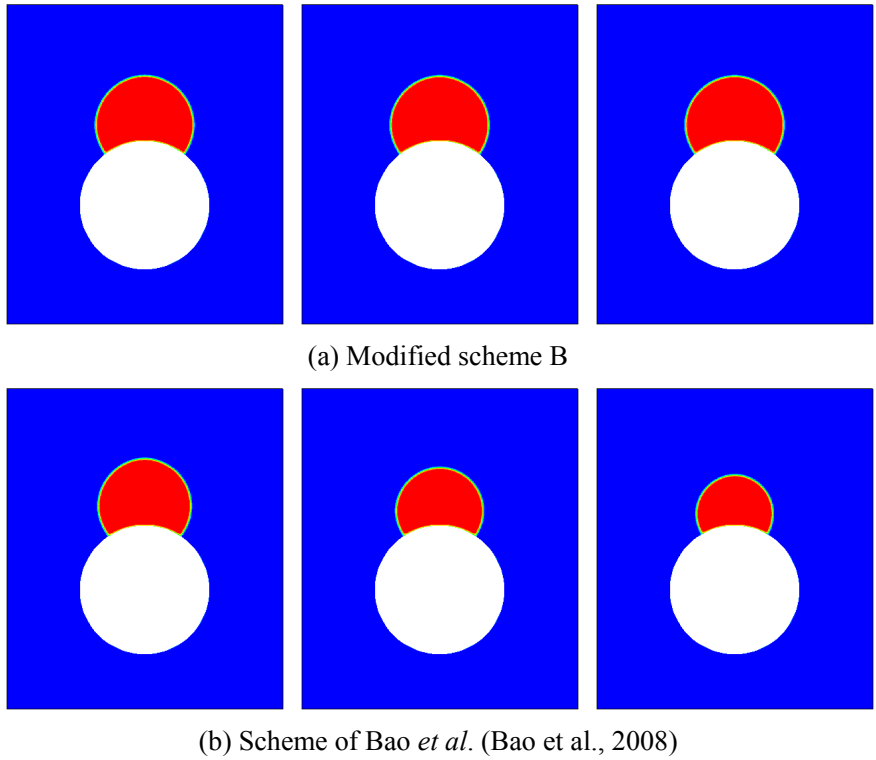

(a) Modified scheme B

(b) Scheme of Bao *et al*. (Bao et al., 2008)

**Fig. 12**. A comparison of the results obtained by the modified curved boundary scheme B and the scheme of Bao *et al*. (Bao et al., 2008). From left to right: $t = 10000\delta_t$, $5 \times 10^4 \delta_t$, $10^5 \delta_t$. The static contact angle is $\theta = 90°$.

## 5. Summary

In this paper, we have investigated the performance of three well-known types of curved boundary schemes in two-phase LB simulations. Through modeling a droplet resting on a circular cylinder, it has been found that, for all the investigated schemes, the simulated droplet rapidly "evaporates" under the non-slip and isothermal conditions, revealing that there exists serious mass leakage at the cylindrical surface due to the curved boundary schemes. The numerical results showed that the MLS boundary scheme yields the smallest mass loss among the investigated curved boundary schemes, while the mass loss caused by the Zhao-Yong scheme, which is a single-node boundary scheme and has the advantage of local computations, is much larger than those of the other schemes.

On the basis of the numerical investigation, we have formulated two modified mass-conservative boundary schemes for curved boundaries based on the MLS boundary scheme. The modified scheme A changes the rest distribution function of the boundary node so as to add the mass leakage back to the system, while the modified scheme B is devised by guaranteeing the balance between the amount of mass carried by the outgoing and incoming distribution functions at each boundary node. Numerical simulations have been carried out to validate the modified schemes. It has been shown that the two modified schemes exhibit better performance than the halfway bounce-back scheme according to the modeling of steady flow past a circular cylinder. Moreover, the simulations of a droplet resting on a circular cylinder have demonstrated that the two modified schemes are capable of conserving mass in two-phase LB simulations.

## Acknowledgments

This work was supported by the National Natural Science Foundation of China (No. 51822606).